\documentstyle[11pt]{article}
\begin{document}
\pagenumbering{arabic}

\def\Bbb{\bf}

\begin{flushright}
\renewcommand{\textfraction}{0}
May 16th 1995\\
hep-th/9505089\\
PEG-04-95\\
\end{flushright}

\begin{center}
{\LARGE {\bf Event-Symmetric Physics} }
\end{center}
\begin{center}
{\Large Phil Gibbs} \\
e-mail to phil@galilee.eurocontrol.fr
\end{center}

\begin{abstract}
I examine various aspects of event-symmetric physics such as
phase changes, symmetry breaking and duality by studying
a number of simple toy-models.
\end{abstract}


Copyright Notice

{ \small This document is Copyright \copyright 1995 by the author Philip E.
Gibbs (phil@galilee.eurocontrol.fr). All rights are reserved. Permission to
use, copy and distribute this document by any means and for any
purpose except profit is hereby granted, provided that both the
above copyright notice and this permission notice appear in all copies.
Reproduction in part or in whole by any means, including, but not
limited to, printing, copying existing prints, publishing by electronic or
other means, implies full agreement to the above non-profit-use clause,
unless upon explicit prior written permission of the author. }

\pagebreak
\begin{center}
{\Large {\bf To Rachel} }
\end{center}
\begin{center}
{\Large (because she is the first to believe it!)}
\end{center}
\pagebreak

\section*{Event-Symmetric Space-Time}

More than anybody else John Wheeler has promoted the idea that physics
must be derived from some deeper pregeometric theory in which space-time
structure arises as an aspect of a more fundamental one
\cite{Whe57-Whe84}. It is very difficult to know how to
approach the task of building such a theory but there have at least been
some worthy attempts which can serve as a source of ideas
\cite{Fin69-Zap95b}.

Many of them take some guiding principle as a basis for constructing a theory.
It may be causality, topology or spin structure for example. But in precious
few cases does symmetry enter as a basic necessity. This is surprising since
symmetry has been the most useful tool in constructing successful theories
of physics this century. The difficulty may simply be that nobody has been
able to see how symmetry can be used in a pregeometric theory.

My own belief is that the symmetry so far discovered in nature is just the
tiny tip of a very large iceberg most of which is hidden beneath a sea of
symmetry breaking. With the pregeometric theory of
{\sl event-symmetric physics} I hope
to unify the symmetry of space-time and internal gauge symmetry into one
huge symmetry. I hope that it may be possible to go even further than this.
Through dualities of the type being studied in string theory it may be
possible to include the permutation symmetry under exchange of identical
particles into the same unified structure. Ultimately we may come to
understand the origins of so much symmetry in terms of some metaphysical
theory of theories ( see e.g. \cite{FrNi91}).

The real domain of the event-symmetric formalism seems to be in string
theory. I have explored the construction of event-symmetric string field
theories elsewhere \cite{Gib95a} and hope to return to it later. In this
article I explore a number of much simpler event-symmetric toy models which
provide some useful insight into the nature of event-symmetric physics.

The theory of Event-Symmetric space-time is a discrete approach to quantum
gravity \cite{Gib94a}.  The exact nature of space-time
in this scheme will only become apparent in the solution. Even the number of
space-time dimensions is not set by the formulation and must by a dynamic
result. It is possible that space-time will preserve a discrete
nature at very small length scales. Quantum mechanics must be reduced to a
minimal form. The objective is to find a statistical or quantum definition of
a partition function which reproduces a unified formulation of known and
hypothesised symmetries in physics and then worry about states, observables
and causality later.

Suppose we seek to formulate a lattice theory of gravity
in which diffeomorphism invariance
takes a simple and explicit discrete form. At first glance it would seem that
only translational invariance can be adequately represented in a discrete form
on a regular lattice.
This overlooks the most natural generalisation of
diffeomorphism invariance in a discrete system.

Diffeomorphism invariance
requires that the action should be symmetric under any differentiable
1-1 mapping on a $D$ dimensional manifold $M_D$. This is
represented by the diffeomorphism group $diff(M_D)$. On a discrete
space we could demand that the action is symmetric under any permutation
of the discrete space-time events ignoring continuity altogether. Generally we
will use the term {\it Event-Symmetric} whenever an action has an invariance
under the Symmetric Group $S({\cal U})$ over a large or infinite
set of ``events'' ${\cal U}$. The symmetric group is the group of all
possible 1-1 mappings on the set of events with function composition as the
group multiplication. The cardinality of events on a manifold of any number
of dimensions is $\aleph_1$. The number of dimensions and the topology of the
manifold is lost in an event-symmetric model since the symmetric groups for
two sets of equal cardinality are isomorphic.

Event-symmetry is larger than the diffeomorphism invariance of continuum
space-time.
\begin{equation}
                 diff(M_D) \subset S(M_D) \simeq S(\aleph_1)
\end{equation}
If a continuum is to be restored then it seems that there must be a mechanism
of spontaneous
symmetry breaking in which event-symmetry is replaced by a residual
diffeomorphism invariance. The mechanism will determine the
number of dimensions of space. It is possible that a model could have several
phases with different numbers of dimensions and may also have an unbroken
event-symmetric phase. Strictly speaking we need to define what is meant by
this type of symmetry breaking. This is difficult since there is no
order parameter which can make a qualitative distinction between a broken
and unbroken phase.

The symmetry breaking picture is not completely satisfactory because it
suggests that one topology is singled out and all others discarded by
the symmetry breaking mechanism but it would be preferred to retain the
possibility of topology change in quantum gravity. It might be more accurate
to say that the event-symmetry is not broken. This may not seem to
correspond to observation but notice that diffeomorphism invariance of
space-time is equally inevident at laboratory scales. Only the Poincare
invariance of space-time is easily seen. This is because transformations of
the metric must be included to make physics symmetric under general coordinate
changes. It is possible that some similar mechanism hides the event-symmetry.
I will continue to use the language of symmetry breaking even if it may
not be strictly correct.

It is possible to make an argument based on topology change that space-time
{\it must} be taken as event-symmetric in Quantum Gravity. Wheeler was the
first to suggest that topology changes might be a feature of quantum
geometrodynamics \cite{Whe57}. Over the past few years the arguments in
favour of topology change in quantum gravity have strengthened see e.g.
\cite{BaBiMaSi95}. If we then ask what is the correct symmetry group in
a theory of quantum gravity under which the action is invariant, we
must answer that it contains the diffeomorphism group $diff(M)$ for any
manifold $M$ which has a permitted topology. Diffeomorphism groups are
very different for different topologies and the only reasonable way to
include $diff(M)$ for all $M$ within one group is to extend the group to
include the symmetric group $S(\aleph_1)$. There appears to be little
other option unless the role of space-time symmetry is to be abandoned
altogether.

There is another theory which would benefit from a formulation in which
fields have non-local interactions independent of distance of the type
postulated in event-symmetric theories. The theory shows that fine-tuning
of the constants of nature could be explained in such circumstances
\cite{BeFrNi95}.

It is unlikely that there would be any way to distinguish a space-time with an
uncountable number of events from space-time with a dense covering of a
countable number of events so it is acceptable to consider models in which the
events can be labelled with positive integers. The symmetry group $S(\aleph_1)$
is replaced with $S(\aleph_0)$. In practice it may be necessary to regularise
to a finite number of events $N$ with an $S(N)$ symmetry and take the large $N$
limit while scaling parameters of the model as functions of $N$.

Having abandoned diffeomorphisms we should ask whether there can remain
any useful meaning of topology on a manifold. A positive answer is
provided by considering discrete differential calculus on sets and finite
groups \cite{DiMu94a}.

In some of the more physically interesting
models the symmetry appears as a sub-group of a larger symmetry such as
the orthogonal group $O(N)$. It is also sufficient that the Alternating
group $A(N)$ be a symmetry of the system since it contains a smaller
symmetric group.
\begin{equation}
                 S(N) \subset A(2N)
\end{equation}

The definition of the term event-symmetric could be relaxed
to include systems with invariance under the action of a group which has
a homomorphism onto $S(N)$. This would include, for example, the braid
group $B(N)$ and, of course, quantum groups such as $SL_q(N)$.

Renormalisation and the continuum limit must also be considered but it is not
clear what is necessary or desired as renormalisation behaviour. In
asymptotically free quantum field theories with a
lattice formulation such as QCD it is normally assumed that a continuum limit
exists where the lattice spacing tends to zero as the renormalisation group
is applied. In string theories, however, the theory is perturbatively finite
and the continuum limit of a discrete model cannot be reached with the
aid of renormalisation. It is possible that it is not necessary to have an
infinite density of events in space-time to have a continuum or there may
be some alternative way to reach it, via a q-deformed non-commutative geometry
for example.

It stretches the imagination to believe that a simple event-symmetric model
could be responsible for the creation of continuum space-time and the
complexity of quantum gravity through symmetry breaking, however, nature has
provided some examples of similar mechanisms which may help us accept the
plausibility of such a claim and provide a physical picture of what is
going on.

Consider the way in which soap bubbles arise from a statistical physics
model of molecular forces. The forces are functions of the relative
positions and orientations of the soap and water molecules. The energy is
a function symmetric in the exchange of any two molecules of the same
kind. The system is consistent with the definition of event-symmetry since it
is invariant under exchange of any two water or soap molecules and therefore
has an $S(N) \otimes S(M)$ symmetry where $N$ and $M$ are the number of water
and soap molecules. Under the right conditions the symmetry breaks
spontaneously to leave a diffeomorphism invariance on a two dimensional
manifold in which area of the bubble surface is minimised.

Events in the soap bubble
model correspond to molecules rather than space-time points. Nevertheless, it
is a perfect mathematical analogy of event-symmetric systems where the
symmetry breaks in the Riemannian sector to leave diffeomorphism invariance in
two dimensions as a residual symmetry. Indeed the model illustrates an
analogy between events in event-symmetric space-time and identical particles
in many-particle systems.
The models considered further are more sophisticated than the molecular
models. However, the analogy between particles and space-time events remains
a useful one.

It might be asked what status this approach affords to events themselves.
Events are presented as fundamental entities almost like
particles. Event orientated models are sometimes known as Whiteheadian
\cite{Whi29} but Wheeler preferred to refer to a space-time viewed as a set
of events without a geometric structure as a ``bucket of dust''
\cite{Whe64,MiThWh73}.
In some of the models we will examine it appears as if events
are quite real, perhaps even detectable. In other models they are more
metaphysical and it is the symmetric group that is more fundamental.
Indeed the group may only arise as a subgroup of a matrix group and the
status of an event is then comparable to that of the component of a vector.
Then again in the discrete string models we will see that events have
the same status as strings.

The concept of event-symmetric space-time fits well into the framework of
non-commutative geometry. It has been shown \cite{CoLo91} that by defining
differential geometry on a space-time consisting of a manifold times a
discrete set of points it is possible to give a geometric interpretation of
theories such as the Electro-Weak Standard Model \cite{Wei67} in which the
Higgs field arises naturally from the generalised connection. If we could
start from a non-commutative geometry defined on just a large discrete
set of points with an event-symmetric formulation then this would make sense
of these model building techniques.

A number of Event-Symmetric models will be described in this paper. Some of
these can best be understood as statistical theories with a partition function
defined for a real positive definite action.
\begin{equation}
                  {\Large Z = \int e^{-S} }
\end{equation}
Others can only be considered as quantum theories for which the action need
not be positive definite provided the partition function is well defined
\begin{equation}
                  {\Large Z = \int e^{iS} }
\end{equation}
It is not always clear when such an integral should be considered well
defined. For example the action,
\begin{equation}
                   S = x^2 - y^2
\end{equation}
gives a well defined quantum partition function in the two variables $(x,y)$
but if the variables are transformed by a 45 degree rotation to $(u,v)$, the
action becomes
\begin{equation}
                   S = 2uv
\end{equation}
for which the integral is not well defined.

It might be safer to consider only positive definite actions and assume that
in a physically valid theory, the only difference between the statistical
event-symmetric model and the quantum one should be a factor of $i$ against
the action in the exponential. We might expect that in the statistical version
the Event-Symmetry will break to give Riemannian space-time with a Euclidean
signature metric while in the quantum version it breaks to give the physical
Einsteinian space-time theory with Lorentzian signature metric.

But is that realistic, after all, continuum Lagrangian densities
for field theories in Minkowski space-time are made non-positive definite by
the signature of the metric? It is not clear what conditions should be placed
on the form of an event-symmetric action to ensure a well defined tachyon free
quantum theory which produces dynamically the correct Lorentz signature. Even
in continuum theories this is an interesting question and it is believed that
a Lorentzian signature is preferred for certain theories in 4 and 6 dimensions.
\cite{Gre92,CaGr93,OdRo94}.


\section*{Event-Symmetric Ising Models}

The simplest event-symmetric model is the event-symmetric Ising model. This
consists of a large number $N$ of feromagnets represented by spin variables
\begin{equation}
                    s_a  = +1 or -1   for   (a = 1,...,N)
\end{equation}
Each spin interacts equally with every other spin according to the action,
\begin{equation}
                    S = \beta \sum_{a<b} s_a s_b
\end{equation}
This has $S(N)$ invariance since it is symmetric in spin permutations
and an additional $Z_2$ invariance under global spin reversal.
Solving this model is not difficult. The partition function is
\begin{equation}
                    Z = \sum_{\{s_a\}} e^{-S}
\end{equation}
Write this as a sum over states with $K$ negative spins
and $N-K$ positive spins.
\begin{equation}
             Z = {\large\sum} C^N_K exp{\large(}(\beta/N)[(N/2)(N-1)
		    - 2 K(N-K)]{\large)}
\end{equation}

In the large $N$ limit this can be written as an integral over a variable
\begin{equation}
                         p = K/N
\end{equation}
\begin{equation}
      Z = {\large \int}_0^1 dp exp \{ N ( \overline{\beta}[1/2 - 2 p(1-p)] -
	  p ln(p) - (1-p)ln(1-p) ) \}
\end{equation}
In this equation we have scaled $\beta$ as a function of $N$ such that
$\overline{\beta} = \beta N$ is kept constant.

The function in the exponential has one minimum at $p = 1/2$ for $\beta \leq 1$
and two minima for $\beta > 1$. The large $N$ limit forces the system into
these minima so there is a second order phase transition
at $\beta = 1$ with the $Z_2$ spin symmetry broken above. The $S(N)$
event-symmetry is not broken in this model.

Although such a model seems quite trivial there is some interest in
generalisations where the $Z_2$ symmetry is replaced with unitary matrix
groups \cite{RoTa94}.

For the gauged version the spins are placed on event links. There are
therefore $(1/2)N(N-1)$ spins
\begin{equation}
                     s_{ab}  =  +1 or -1 ,   a < b
\end{equation}
And the action is now a sum over triangles formed from three links
\begin{equation}
                    S = \beta \sum_{a<b<c}    s_{ab} s_{bc} s_{ac}
\end{equation}
This model again has an $S(N)$ event-symmetry but the $Z_2$ symmetry is
now a gauge symmetry. This is already too complicated to solve exactly
by any obvious means.

The most interesting thing that can be said about this model is that it
is dual to a model of surfaces which can be compared to string world
sheets. Let $T$ be the set of all possible triangles with vertices in the
set of events. i.e.,
\begin{equation}
                    T = {(a,b,c) : a < b < c}
\end{equation}
then,
\begin{equation}
   Z = {\large\sum}_{\{s_{ab}\}} \prod_{(a,b,c) \in T}
     {\Large(} cosh\beta + s_{ab} s_{bc} s_{ac} sinh\beta {\Large)}
\end{equation}
\begin{equation}
     = cosh\beta^{1/6 N (N-1) (N-2)} {\Large\sum}_{R \in 2^T}
        tanh\beta^{|R|} {\large\sum}_{\{s_{ab}\}} \prod_{(a,b,c) \in R}
	s_{ab} s_{bc} s_{ac}
\end{equation}
The inner sum over the product is zero except when the subset $R$ of triangles
contains each link variable an even number of times. Such a subset can be
considered a surface formed from the triangles. It may be made up of
several pieces and it may cross itself at links. The outer sum can then
be replaced with a sum over surfaces $B$ and if the number of
triangles in $B$ is interpreted as its area $A(B)$ then an effective action
is left given by
\begin{equation}
                    S' = ln(tanh\beta) A(B)
\end{equation}
This is analogous to the Area action for a first quantised bosonic string but
is defined on an event-symmetric lattice instead of a $D$ dimensional
continuous target space.


\section*{Molecular Models}

Insight into event-symmetric statistical physics models and the possibilities
for symmetry breaking can be gained from molecular models. In general a
molecular model describes a large number $N$ of molecules given by
their position ector $X_a$ and orientation matrices $O_a$ in a $D$ dimensional
Euclidean hyperspace. For simplicity kinetic energy is discarded and the
interactions are described by an energy potential,
\begin{equation}
                  E = V( \{X_a\},\{O_a\} )
\end{equation}
The potential should tend rapidly to a constant at large distances in order
to suppress long range interactions,
and should by invariant under global translations and rotations.
Furthermore the potential should be invariant under exchange of any
two molecules so that the description event-symmetric can be justified.
An analogy then exists between the molecular model and a model of
an event-symmetric space-time in which events correspond to molecules.

The simplest possibility is to model space-time as a critical solid
\cite{Orl93}. For a suitable action symmetric in exchange of molecules
they can model a critical solid at a second order melting phase transition.
This gives rise dynamically to what
might be interpreted as a $D$ dimensional curved manifold. In this case
the number of dimensions is predetermined and it is difficult to see how the
space-time could form different topologies. The event-symmetry is broken
in the solid phase since the molecules settle into a lattice configuration
leaving a residual translation symmetry. Near the phase transition a
scaling behaviour might be observed with a larger residual symmetry
in the critical limit. In a gas phase the model would be fully
event-symmetric.

More persuasive models might be constructed by attempting to simulate
a molecular model of soap film bubbles. A single species of molecule in
$D$-dimensional hyperspace with an
orientation dependent energy potential favouring alignment should
be sufficient. In such a model the molecules would tend to form
lattices on lower $d$-dimensional hyper-surfaces at low temperatures.

To make this more concrete we shall look for a suitable energy
function. Take it to be a sum of potential energies between
molecule pairs in $D = 3$ space.
\begin{equation}
                  S = \beta E = \sum_{ab} V(X_a,X_b,O_a,O_b)
\end{equation}
Spatial symmetry is ensured if the potential is a function of
the following scalar invariants,
\begin{equation}
                  r_{ab} = |X_a - X_b|
\end{equation}
\begin{equation}
 cos(\theta_{ab}) = \hat{i} . O_a (X_a - X_b) / r_{ab},
             0 \leq \theta_{ab} \leq \pi
\end{equation}
Where $\hat{i}$ is a unit vector in the axis of the molecule.
A likely looking possibility in terms of these invariants is,
\begin{equation}
 S = \beta \sum_{ab} [ 2 r_{ab}^{-1} - sin^2(\theta_{ab}) ] e^{-r_{ab}}
\end{equation}
The minimum energy configuration for two molecules is when they are a
distance $r = 1 + \sqrt 3 $ apart and are oriented perpendicular
$\theta_{ab} = \pi/2$ to the
line which joins them. With many particles, the minimum energy state is
achieved when they are packed into a triangular lattice in a $d = 2$ plane. The
spacing can be computed numerically to be $r = 2.13$. The configuration
is stable against movement of a molecule out of the plane.

At low temperatures (large $\beta$) the molecules would stay near the
lattice positions so there would be a solid crystalline phase
At higher temperature the molecules might be expected to flow in the
plane simulating a liquid phase. There should be a phase transition
between the solid and liquid phase at which scaling behaviour might
be observed. At higher temperatures the bubble would evaporate
into a gas phase and the full event-symmetry would be restored.

The most interesting part of the phase diagram might be the transition from
the crystalline to fluid phase. Similar transitions have been studied
numerically in the context of lattice gravity where there might be an
interesting transition between fixed triangulations and random triangulations
on surfaces \cite{BaJo93}.

The benefit of bubble models is that surfaces with different
topologies are possible and diffeomorphism invariance is a
possibility as a residual symmetry. It is probable that the $D = 3, d = 2$
model can be generalised to higher dimensions.

The analogy between statistical molecular models forming soap films and
event-symmetric models forming space-time can be a very useful one to
help us visualise the physics of event-symmetric theories. We can really
imagine space-time evaporating at high temperature for example.
There are, however, some
important differences which must be born in mind: A molecular model is
formed as an embedding in a higher dimensional space whereas the most
interesting event-symmetric models are defined in some kind of Machian
void; In molecular models the discrete objects are always
hard objects which can be detected individually whereas in event-symmetric
models events may be only a bases of an algebra with no existence as
individual objects. Finally the molecular models are models in statistical
physics with no time evolution whereas a physical event-symmetric model
must be a full quantum theory even if time is not an exact concept.

Molecular models are well understood in terms of equilibrium thermodynamics
under changes of temperature and pressure. It may be possible to define
a phase diagram of event-symmetric theories in a similar way. It would
then be possible to think of the formation of space-time as a condensation
process. If space-time behaves like molecular models then it may be
possible to go from the broken phase to the unbroken phase of gravity
just as it
is possible to go from a gas to a liquid at high pressure without passing
through a phase transition. There might also be physical significance
of critical points. Of course it might not be possible to define
temperature and pressure in an event-symmetric model.


\section*{Symmetric Random Graph Models}

Since we are looking for some kind of spontaneous symmetry breaking in which
the number of dimensions is dynamically determined
it makes sense to investigate systems on which we can attempt to define
dimensionality. The simplest such structure would be a random graph in
which $N$ nodes are randomly pairwise connected by up to $1/2 N (N-1)$ links
\cite{DaPi79,Bol85,CiGo89,Ant94a,Req95}. An event-symmetric action is a
function of the connections which is
invariant under any permutation of nodes. For example, actions defined
as functions of the total number of links and the total number of
triangles in a graph would be event-symmetric.

The principle is that on a graph we can define dimensionality from its
connectivity. For a given node we can define a function $L(s)$, the
number of nodes which can be reached by taking at most $s$ steps
along links. If $L(s)$ has a power law on an infinite graph,
\begin{equation}
           L(s) \rightarrow s^D as  s \rightarrow \infty
\end{equation}
then the graph has dimension D. It may also be possible to determine
dimensionality from topology of a finite graph \cite{Eva94} or, if the
links are bidirectional, the topology can be derived from an analysis of
posets \cite{DiMu94b}.

If a suitable mechanism of symmetry breaking is effected on the system
the graphs generated statistically from the action may have some
finite dimension. The number of dimensions could differ from one
phase of the system to another. There could also be phases in which
the event-symmetry is unbroken and the number of dimensions can be
considered infinite.

The random graph models are similar in some ways to the random lattice
models of quantum gravity but are much simpler since there is no need
to apply constraints which select the topology in the formulation.
Instead the sum is over all discrete topologies. It is also much easier
to ensure that an action is positive definite.

Wheeler was one of the first to think about this sort of space-time
model \cite{MiThWh73}. He Likened a random graph to a sewing
machine stitching together a space-time. Wheeler found that such models
did not appeal to his taste in simplicity. Nevertheless they are a
useful starting point for exploration of event-symmetric space-times
even if they are unphysical.

Since there are no other symmetries to guide our choice of action we
might consider heuristic criteria to contrive an action which might
exhibit spontaneous symmetry breaking of the event-symmetry. As a first guess
it might be reasonable to consider an action which favours triangles
but disfavours links. The action can be written in terms of link
variables  $l_{ab}$
\begin{equation}
            l_{ab} = 1 if the nodes a and b are linked, = 0 otherwise
\end{equation}
Define
\begin{equation}
                       V_a = \sum_b l_{ab}
\end{equation}
\begin{equation}
                       T_a = \sum_{b,c} l_{ab} l_{bc} l_{ac}
\end{equation}
I.e. $V_a$ is the valence of node $a$ and $T_a$ is the number of
triangles in the graph which have a vertex at $a$.
\begin{equation}
             S = - \sum_a [(\beta/N^2) T_a  -  (\alpha/N^2) V_a^2 ]
\end{equation}
A simple mean field analysis can be performed where each link is connected
with a probability $p$. Then
\begin{equation}
               T_a  = N^2 p^3
\end{equation}
\begin{equation}
	       V_a  = N p
\end{equation}

Taking into account that the number density of states as a function
of p this gives an effective action of,
\begin{equation}
           S =  N[ p ln(p) + (1-p) ln(1-p) - \beta p^3 + \alpha p^2 ]
\end{equation}

This suggests a phase transition along approximately $\beta/\alpha = 1$
with p close to one for $\beta > \alpha$ and p close to zero for
$\beta < \alpha$

Further mean field analysis of this model and other similar models is possible.
An extension to the treatment given here would be to consider a mean field
analysis of the situation where the graph breaks down into small isolated
parts. Linkage between nodes within each part can be given a probability
$p$ while linkage between nodes in different parts can be given a
probability $q$. A mean field analysis for a particular Event-Symmetric
action might suggest that an asymmetric phase existed with $q$ small and $p$
close to one. It is possible that this could be taken as a signal that
other forms of Symmetry Breaking were a possibility for that action

Numerical simulations could also be used to look for evidence of
Event-Symmetry breaking. It may be possible to construct models
in this way which have residual structures with finite dimensional
symmetries.

In fact there is one very simple random graph model in which the symmetry
breaks to one dimension. This is given by the action,
\begin{equation}
                  S = \beta \sum_a (V_a - 2)^2
\end{equation}
At high $\beta$ the model forces exactly two links to meet at each vertex of
the graph. I.e. it must break down into rings which can be considered as one
dimensional spaces.


\section*{Dynamical Triangulations}

Lattice studies of pure gravity start from the Regge Calculus
\cite{Reg61} in which space-time is ``triangulated'' into a simplicial complex.
The dynamical variables are the edge lengths of the simplices.
In 4 dimensions an action which reduces to the usual Einstein Hilbert action
in the
continuum limit can be defined as a sum over hinges in terms of facet areas
$A_h$ and deficit angles $\delta_h$ which can be expressed in terms of the edge
lengths.
\begin{equation}
                  S = \sum_h k A_h \delta_h
\end{equation}
The model can be studied as a quantised system and
this approach has had some limited success in numerical studies
\cite{RoWi81,RoWi84,Ham85,Ham91}.

A variation of the Regge calculus is to use dynamical triangualtions
An action with fixed edge
lengths but random triangulations \cite{BoKaKoMi86} is given by,
\begin{equation}
                 S = - \kappa_4 N_4 + \kappa_0 N_0
\end{equation}
The partition function is formed from a sum over all possible triangulations
of the four-sphere. $N_4$ is the number of four simplices in the triangulation
and $N_0$ is the number of vertices. The constant $\kappa_4$ is essentially
the cosmological constant while $\kappa_0$ is the gravitational coupling
constant. Random triangulations
of space-time appear to work somewhat better than the Regge Calculus with
a fixed triangulation.

It is possible to construct event-symmetric models which reduce to
dynamical triangulations of manifolds as limiting cases. This is certainly
an interesting prospect given the provisional success of dynamical
triangulations as models of Riemannian sector quantum gravity in numerical
simulations.

Two dimensional manifolds can be broken down into triangles so we define
a triangle variable $t_{abc}$ which takes the value $1$ or $0$ according
to whether or not the three events $a$, $b$ and $c$ are the vertices of
a triangle in the triangulation. In an event-symmetric model these are
simply dynamic variables defined for any three events analogous to the
link variables of the random graph models.
The following constraints are applied,
\begin{equation}
                     t_{abc} = t_{bac} = t_{acb}
\end{equation}
\begin{equation}
		     t_{aac} = 0
\end{equation}
The number of triangles meeting at an edge defined by two events $a,b$ is
\begin{equation}
                     L_{ab} = \sum_c t_{abc}
\end{equation}
In a dynamical triangulation of a manifold this must be everywhere either
zero or two. We can define an action,
\begin{equation}
                 S = \beta \sum_{ab} L_{ab}^2 (L_{ab} - 2)^2
\end{equation}
In the high $\beta$ limit this forces the triangles to form a triangulation
of some manifold but there is nothing to ensure that the manifold is
connected or that it has to be oriented. The sum must be over all
topologies.

It is possible to force the manifolds to be oriented. This can be done
by allowing the value for $t_{abc}$ to be $0$, $1$ or $-1$, with the
constraints
\begin{equation}
                     t_{abc} = -t_{bac} = -t_{acb}
\end{equation}
An edge in the triangulation must have a contribution from a negative triangle
and a positive triangle to ensure the surface is oriented. A suitable
action can also be contrived for this case. It is important to make a
restriction to oriented manifolds since otherwise parity violation would
not be possible.

Such models can easily be generalised to give simplicial decompositions of
higher $D$ dimensional manifolds by defining variables with D indices which
are fully antisymmetric. This justifies the assertion that dynamical
triangulations are limiting cases of event-symmetric systems provided a
sum over all topologies is included.

These are very crude models but there are a couple of important lessons
to be learnt here. The first is that event-symmetric models which
incorporate higher dimensional objects than the simple events and links
which appeared in the random graph models seem to have better potential
for forming space-time structure. There are intersting models with field
variables defined on simplex like structures of arbitrarily high dimension
which might be very promising in this respect.

The second lesson is that continuum space-time models of quantum gravity
which include a suitable weighted sum over topologies can be seen as
limiting cases of event-symmetric models. Heuristically we might conclude
that the sum over topologies factors out the diffeomorphic structure
of theories with diffeomorphism invariance on manifolds leaving a completely
event-symmetric theory. This may be physically important if 4 dimensional
quantum gravity at low energy can thus be seen as a limit of a more complete
event-symmetric theory in which space-time dimension is not precisely defined.


\section*{Event-Symmetric field Theory}

The random graphs are interesting as models of space-time but ultimately
we are interested in modelling field theories. It is conceivable that field
theories with continuous variables could somehow arise out of theories with
discrete variables but if we are to see gauge symmetries of the type found in
Yang-Mills Theories represented in an exact discrete form at a more fundamental
level then continuous variables must be used.

The simplest event-symmetric field theory would be given by a scalar field
$\phi_a$ defined on events $a$. We might define an action of the form,
\begin{equation}
                 S = \sum_a m \phi_a^2 + \sum_{a<b} (\phi_a - \phi_b)^2
		        + \sum_a g \phi_a^n
\end{equation}
If $n>2$ is even and $g$ is positive then a statistical field theory can be
defined with partition function,
\begin{equation}
                  Z = \int exp(-S) d^N \phi
\end{equation}
For odd $n$ the action is not positive definite but a quantum field theory
is well defined with the partition function,
\begin{equation}
                  Z = \int exp(iS) d^N \phi
\end{equation}
It is useful to look at the perturbation theory of such models. First the
non-interacting $g = 0$ case should be solved. The quadratic part of the
action takes the form,
\begin{equation}
                 S = \sum_{a,b} E(m + (N-1),-1)_{ab}\phi_a\phi_b
\end{equation}
where the notation $E(d,e)$ is used to denote an {\it event-symmetric matrix}
which has the value $d$ for each diagonal element and the value $e$ for each
off diagonal element. The propagator will be given by the inverse of this
matrix which is easily found to be,
\begin{equation}
                 E(m + (N-1),-1)^{-1} = E(m+1,1)/[m(m+N)]
\end{equation}
This would be singular in the case where $m = 0$.

The propagators in the interacting case can be represented as a sum over
Feynman diagrams which take the form of fixed valence graphs. I.e. the
graphs have exactly $n$ edges joined at each vertex. For each vertex there
is a sum over events and a factor $g$. For each edge there is a propagator
factor $1/[m(m+N)]$. If we take $m$ to be small and ignore accidental
symmetry factors then the contribution of each graph is,
\begin{equation}
                 (gN)^{N_0} (mN)^{-N_1}
\end{equation}
Where $N_0$ is the number of nodes and $N_1$ is the number of edges.
\begin{equation}
                 2 N_1 = n N_0
\end{equation}
A sum over fixed valence graphs can be considered as another type of
event-symmetric random graph model. This simple type can be solved.
There are more interesting versions constructed from Ising models
on the nodes of fixed valence graphs which can also be studied analytically
by relating them to the perturbation theory of event-symmetric scalar
field theories \cite{BaCaPe94,Joh94,DoKu92}. From our point of view gauged
Ising models on fixed valence random graphs might be even more interesting.

If the ultimate aim is to produce event-symmetric models of real physics
then it will be necessary to introduce further symmetries such as gauge
symmetry. The Event-symmetric Ising gauge model can be combined with a
random graph model giving a model with link variables which can take
three values -1, 0 or +1. Such models are interesting to study for
event-symmetry breaking because the duality transformation can still
be applied to give a dual model of strings on a random graph.

To go further the $Z_2$ gauge symmetry can be extended to gauge symmetry
of other groups such as $U(1)$, $SU(3)$ etc. The link variables then takes
values zero or an element of the group. Such models represent a kind of
gauge glass \cite{BeBrNi87,FrNi91}. We shall see that there
are more unified ways to combine event-symmetry and gauge symmetries.

For the symmetry to break in the way we desire, i.e.
leaving a finite dimensional topology, the events
will have to organise themselves into some arrangement where there is an
approximate concept of distance between them perhaps defined by correlations
between field variables. Matrix elements linking events which are
separated by large distances would have to be correspondingly small. Only
variables which are localised with respect to the distance could have
significant values.

Field theory can be extended further than placing field variables on just
sites and links between sites. They can also be attached to higher dimensional
cells or simplices such as triangles and tetrahedrons. This can be understood
as the field theory extension of dynamical triangulations. It may be easier
to analyse dimensional phase transitions in such a context. Useful work in
this context had been produced by Jourjine and Vanderseypen who express field
theory on cell complexes in the mathematical language of homology and
cochains \cite{Jou85,Jou86,Jou87,Van93}.


\section*{Random Matrix Models}

An important class of event-symmetric model places field variables $A_{ab}$
on links joining all pairs of events $(a,b)$. A suitable action must be a
real scalar function of these variables which is invariant under exchange
of any two events.

The link variables $A_{ab}$  can be regarded as the elements of a matrix $A$.
If the direction of the link is irrelevant the matrix can be taken to be
either symmetric or anti-symmetric. If there are no self links the diagonal
terms are zero so it is natural to make the matrix anti-symmetric.
\begin{equation}
               A_{ab} = - A_{ba}
\end{equation}
A possible four link loop action is
\begin{equation}
               S = \beta \sum_{a,b,c,d} A_{ab} A_{bc} A_{cd} A_{da} +
	           m \sum_{a,b} A_{ab} A_{ab}
\end{equation}
\begin{equation}
                 = \beta Tr(A^4) + m Tr(A^2)
\end{equation}
which is an invariant under $O(N)$ similarity transformations on the matrix.

The symmetric group $S(N)$ is incorporated as a sub-group of $O(N)$
represented by matrices with a single one in each row or column and all
other elements zero in such a way that the matrix permutes the elements
of any vector it multiplies. This suggests that in general we should consider
actions which are functions of the traces of powers of the matrix $A$.
The same idea can be extended to unitary groups by using complex variables for
hermitian matrices or symplectic groups by using quarternions.

This is an appealing scheme since it naturally unifies the $S(N)$ symmetry,
which we regard as an extension of diffeomorphism invariance, with gauge
symmetries. If the symmetry broke in some miraculous fashion then it
is conceivable that the residual symmetry could describe quantised gauge
fields on a quantised geometry.

Consider for example a discrete gauge $SO(10)$ symmetry on a hypercubic lattice
of $N = M^4$ points. The full symmetry group $Lat(SO(10),M)$ is generated
by the gauge group $SO(10)^N$ and the lattice translation and rotation
operators. A matrix representation  of this group in $10N \times 10N$
orthogonal
matrices can be constructed from the action of the group on a 10 component
field placed on lattice sites. The group is therefore (isomorphic to) a
sub-group of an orthogonal group.
\begin{equation}
               Lat(SO(10),M) \subset O(10 N)
\end{equation}

We can imagine a mechanism by which the $O(N)$ symmetry of a matrix model
broke to leave a residual $Lat(SO(10),M)$ symmetry.
It seems highly unlikely, however, that such an exact form of symmetry
breaking could arise spontaneously.

This type of random matrix model has been extensively studied in the
context where $N$ is interpreted as the number of colours or flavours. (see
\cite{ItDr89,Kak91} ) The event-symmetric paradism suggests an alternative
interpretation in which $N$ is the number of events times the number of
colours \cite{KaWe85}.

This unification of space-time and internal gauge symmetries might be
compared with the similar achievement of Kaluza-Klein theories in which
the symmetry is also extended and assumed broken. Here the symmetry is
much larger and could be compared with a Kaluza-Klein theory which had
an extra dimension for each field variable \cite{KaSu83}.

One interesting result for matrix models which is responsible for them
attracting so much interest is that the perturbation theory of
an $SU(N)$ matrix model in the large $N$ double scaling limit is equivalent
to two dimensional gravity or a $c=0$ string theory \cite{tHo74,Kaz89}.

To see this observe that the Feynman diagrams form graphs with nodes of valence
$v$ corresponding to terms in the action given by the trace of matrices to
the power $v$. The edges meeting at a given node are cyclically ordered in
correspondence to the multiplication order of the matrices in the trace.
Given this ordering it is possible to form a surface from the graph. Faces
formed from edges are identified by following loops of edges round the graph
in such a way that the next edge in the loop is consistent with the cyclic
ordering of the edges at each node.

The diagrams are thus in one to one correspondence with facetting of
surfaces with restricted vertex valency. The sum over diagrams for surfaces
of any given topology defines a field theory on the phase space of facet
decompositions of that surface. It is found that the contributions from
a given diagram is in fact a topological invariant of the surface. This
universality is explained by a correspondence between primitive moves changing
the decomposition and the fact that the matrix algebra is associative and
semi-simple \cite{FuHoKa92}.

Many generalisation to multi-matrix models have been studied.
In describing the general forms for actions that we can allow for these
models we must apply a certain locality principle as well as the gauge
invariance. The action must be restricted to forms in which it is the sum of
terms which are functions of the trace of matrix expressions and which do not
separate into products of two or more such scalar quantities. For example
if there are two matrices A and B defining the field variables then
the action could contain terms such as,
\begin{equation}
                        tr(ABAB)
\end{equation}
but not,
\begin{equation}
                       tr(AB)^2
\end{equation}
or
\begin{equation}
                       tr(A)tr(B)
\end{equation}
This locality condition is important when selecting suitable actions for
models which might exhibit dimensional symmetry breaking, since otherwise
the broken phase would have long range interactions.


\section*{Random Tensor Models}

The matrix models have several possible generalisations to tensor models and
models with fermions. In each case the action can be a function of any set of
scalars derived from the tensors by contraction over indices, with the indices
ranging over space-time events.

In tensor models it is often useful to associate tensors which have certain
symmetry constraints with geometric objects having the same symmetry in
such a way that the indices correspond to vertices of the object. For example
a rank 3 tensor which is symmetric under cyclic permutations of indices
\begin{equation}
                       T_{abc} = T_{bca}
\end{equation}
can be associated with a triangle joining the three vertices $a$, $b$
and $c$. Often models of interest use fully anti-symmetric rank-$d$
tensors which can be associated with an oriented $d$-simplex.

If symmetry breaking is going to separate events then locality is important.
Happily there is a sense in which we can define local interactions
independently of any symmetry breaking mechanisms within the general context
of tensor models.

In each of the models we have looked at there are field variables which have
an association with one or more events. In matrix models the matrix elements
$A_{ab}$ are associated with two events indexed by $a$ and $b$. They represent
an amplitude for the connection of those two events as linked neighbours in
space-time. In tensor
models a tensor of rank $r$ is likewise associated with $r$ events. When
symmetry breaking occurs we expect the events to somehow spread themselves
over a manifold. A field variable associated with events which are not
neighbours on the manifold should be physically insignificant, this will
usually mean that it is very small. Field variables which are associated
with a local cluster of events can be large and are significant in the
continuum limit. Two such field variables which are localised around
different parts of the manifold should not be strongly correlated. They
must therefore not appear in the same interaction term of the action
unless multiplied by some other small field variable.

This heuristic picture leads to a definition of locality in which
interaction terms are excluded if they separate into the product of
two parts which do not share events. More precisely we can define an
{ \it interaction
graph}  corresponding to any interaction term which has a node for each
variable in the term. Two nodes are linked if the variables are
associated with at least one event in common. We then say that the
model satisfies the {\it weak locality principle} if all interaction
graphs are connected. We will also say that it satisfies
the { \it strong locality principle} if every pair of nodes is
connected in all graphs. I.e. they are triangles, tetrahedrons or
higher dimensional simplexes.

As an example, a matrix model with terms given by the traces of powers
of the matrix,
\begin{equation}
                  I_n = tr(A^n)
\end{equation}
are weakly local because the graphs are $n$-sided polygons with possibly
other links. If the model includes only powers up to the third then it
is strongly local.

It is reasonable to expect that physical event-symmetric field theories
would have to be at least weakly local. There seems to be no special
reason to demand that a theory should be strongly local but it is
notable that this condition often reduces the number of possible interaction
terms from infinity down to one or two without seeming to exclude the
most interesting models.

There have also been interesting studies based on rank three tensors where
the perturbation theory describes the joining of tetrahedral simplices
to build a three dimensional space \cite{AmDuJo91,Sas91,GoGr91}. However,
these models do not exhibit the same universality properties that make the
matrix models so powerful. This fault has been corrected by Boulatov who
replaces tensors with functions on groups and defines an action which
generates 3 dimensional lattice models \cite{Bou92a,Bou92b,Bou93}.


\section*{Supersymmetric Models}

It would be an obvious next step to generalise to supersymmetric matrix models
\cite{MaPa90,GiPe91,AlMa91,Yos91}. So far we have matrix models based a
families of groups such as $O(N)$,
$SU(N)$ or $Sp(N)$. Tensor representations and invariants can be used to
construct models with commuting variables, anticommuting variables or
both. Similarly we can define models based on supermatrix groups of
which there are also several families such as $SU(L|K)$ and $OSp(L|K)$.
For analysis and classification of supergroups see \cite{Cor89}.

One simple super event-symmetric model has an anti-hermitian matrix
$A$ of commuting variables
\begin{equation}
                   \overline{A}_{ab} = -A_{ba}
\end{equation}
and a vector $\psi$ of anti-commuting variables.
A suitable action could be,
\begin{equation}
 S = m(2 i \overline{\psi}_a \psi_a + A_{ab} A_{ba})
\end{equation}
\begin{equation}
     + \beta( 3 \overline{\psi}_a A_{ab} \psi_b -
        i A_{ab}A_{bc}A_{ca} )
\end{equation}
As well as $U(N)$ invariance this is invariant under a super-symmetry
transform with an infinitesimal anticommuting parameter $\epsilon_b$,
\begin{equation}
 \delta A_{ab} = \overline{\epsilon}_b \psi_a - \epsilon_a \overline{\psi}_b
\end{equation}
\begin{equation}
	  \delta \psi_a = i \epsilon_b A_{ab}
\end{equation}
\begin{equation}
	  \delta \overline{\psi}_a = i \overline{\epsilon}_b A_{ba}
\end{equation}
It is necessary to confess that this model is flawed because the
super-symmetry is not closed. It can be completed with the inclusion
of a single scalar variable but this spoils its locality.

A more general class of models can be constructed from superhermitian
matrices which take a black diagonal form,
\begin{equation}
                 S = \matrix{ A & B \cr iB^\dagger & C \cr }
\end{equation}
where $A$ is a hermitian $K by K$ matrix of commuting variables,
$B$ is a $K by L$ matrix of anticommuting variables and $C$ is a hermitian
$L by L$ matrix of commuting variables. The supertrace is defined
as
\begin{equation}
                 sTr(S) = Tr(A) - Tr(C)
\end{equation}
Actions defined with terms expressed as the supertrace of powers of
the matrices are invariant under a $U(K|L)$ super-symmetry. This can
be interpreted as an event-symmetric model with two types of event since
the supergroup has a sub-group isomorphic to $S(K) \otimes S(L)$. If
there is also a vector with components on events in the model then it would
have commuting variables on one type of event and anticommuting on another.
It is possible to interpret this as an indication that events themselves
have either bosonic or fermionic statistics in this model.

It is encouraging that supersymmetric generalisations of matrix models
can be so easily
constructed on event-symmetric space-time. Demanding supersymmetry helps
reduce our choice of actions
but not actually very much. There are still many different possibilities
like the above which can be constructed from contractions over tensor
representations of supersymmetry groups. With such models we would hope
to find examples of symmetry breaking where the residual symmetry included
space-time supersymmetry but these models are special cases of
matrix or tensor models so they will not be more successful
as a scheme for dimensional symmetry breaking.


\section*{Spinor Models}

If it is not possible to break event-symmetry with simple tensor models
then it is necessary to investigate models with spinor representations
or models with tensors of unlimited rank.

The advantage of spinors is that the dimension of the representations
increases exponentially with $N$. For a model using a finite number of
tensor representations the dimension is only polynomial in $N$.

A simple model would have an $O(N)$ symmetry and a Dirac spinor $\Psi$
representation with $2^{N/2}$ anticommuting components. An invariant
action can be constructed using the gamma matrices in the spirit of
a Gross-Neveu model \cite{GrNe74}.
\begin{equation}
     S = im\overline{\Psi}\Psi + \beta\overline{\Psi}\Gamma_a\Psi
     \overline{\Psi}\Gamma_a\Psi
\end{equation}

This model can be solved by introducing a bosonic variable $\sigma_a$ to
remove the 4th degree term

\begin{equation}
 S = im\overline{\Psi}\Psi + 2\beta\overline{\Psi}\Gamma_a\Psi \sigma_a -
     \beta\sigma_a\sigma_a + (N/2) ln(2 \pi \beta)
\end{equation}

The fermionic variables can then be integrated giving the determinant of
a matrix whose eigenvalues are easily derived.
\begin{equation}
  Z  = (2 \pi \beta)^{N/2}{\LARGE\int} d\sigma^N {\LARGE|} i m I + 2 \beta
   \Gamma_a \sigma_a{\LARGE|} exp( -\beta \underline{\sigma}^2 )
\end{equation}
\begin{equation}
  Z  = (2 \pi \beta)^{N/2}{\LARGE\int} d\sigma^N
     ( 4 \beta^2 \underline{\sigma}^2 +
            m^2 )^M exp( -\beta \underline{\sigma}^2 )
\end{equation}
\begin{equation}
                  M = 2^{N/2-1}
\end{equation}
which can be reduced to an integral over one variable,
\begin{equation}
   Z  = \beta^{N/2} \Gamma(N/2)^{-1}{\LARGE\int_0^\infty} d\sigma ( 4 \beta^2
       \sigma^2 + m^2 )^M \sigma^{N-1} exp( -\beta \sigma^2 )
\end{equation}

By integrating completely we destroy any possibility of symmetry breaking.
It is necessary to introduce some kind of symmetry breaking term and
rework. There are various terms which could be considered but the simplest
is a vector term
\begin{equation}
      S_1 = \overline{\Psi}\Gamma_a\Psi v_a
\end{equation}
By $O(N)$ invariance a vector term can be rotated to have just one component.
So add a term to the action of the form
\begin{equation}
      S_1 = \overline{\Psi}\Gamma_1\Psi v
\end{equation}
then,
\begin{equation}
       Z  = \beta^{N/2} \Gamma((N-1)/2)^{-1}
       {\LARGE\int_0^\infty} d\sigma
       {\LARGE\int_{-\infty}^\infty} d\sigma_1
       ( 4 \beta^2 \sigma^2 + (2 \beta \sigma_1 + v ) ^2 + m^2 )^M
       \sigma^{N-2} exp[ -\beta (\sigma^2 + \sigma_1^2) ]
\end{equation}
The integrand has two maxima in $\sigma_1$ which dominate the integral.
The asymmetry introduced by the vector term causes a shift from one
maxima to the other and dynamically breaks the symmetry. Taking the
limit $a \rightarrow 0$ indicates that spontaneous breaking of
symmetry can arise.

The result is symmetry breaking from $O(N)$ to $O(N-1)$. Although this is
far from being what we are looking for, a mechanism which selects
one event would be interesting if that event could be identified as the
initial event of the universe! c.f. \cite{Mof92,HaLaLy93}.

A better understanding of this kind of model can be found from a different
version in terms of staggered fermions \cite{Sus77} on a $2^N$ lattice, i.e. an
$N$-dimensional hypercube . A
real or complex fermionic variable $\psi_i$ is placed on each lattice site
$i$ and interactions are described in terms of matrices of sign factors
$\Gamma^{ij}_a$ linking edges of the $N$ dimensional hypercube. Where $(i,j)$
is not an edge of the hypercube in direction $a$ the matrix component os zero.
Elsewhere the matrices are taken equal to $\pm 1$ on each edge such that the
product round any square plaquette is $-1$ and such that the matrices are
antisymmetric.

There are many ways to fulfill this but they are all equivalent under some
transformation of sign changes on the fermionic variables. The matrices
satisfy the usual anticommutation relations for Dirac's gamma matrices in
Euclidean $N$-dimensional space.

The action is
\begin{equation}
                 S = im \overline{\psi_i} \psi_i +
		 \beta \psi_i \Gamma^{ij}_a \psi_j
		       \overline{\psi_i} \Gamma^{ij}_a \overline{\psi_j}
\end{equation}

Despite being formulated on a lattice this has an exact $SO(N)$ invariance
which can be seen explicitly by reducing the representation to a family
of antisymmetric tensors
\begin{equation}
                   ( \alpha, \alpha_a, \alpha_{ab}, \alpha_{abc}, \ldots )
\end{equation}
The scalar $\alpha$ is placed on one corner of the lattice. The vector
components are placed on the $N$ sites which are linked to that corner
according to the direction of the link. In general the components of
the rank-$r$ tensor are placed on the $C^N_r$ sites which can be reached
through $r$ links from the corner. With a suitable choice of the sign
factors in the gamma matrices we get
\begin{equation}
      \psi_i \Gamma^{ij}_a \psi_j = \alpha \alpha_a +
      \alpha_b \alpha_{ba} +  \alpha_{bc} \alpha_{bca} + \ldots
\end{equation}
This makes the $SO(N)$ invariance explicit but since the corner was an
arbitrary choice the symmetry should be larger. In fact the model has
a $Spin(N) \otimes Spin(N)$ invariance.

The tensor formulation is interesting because it allows us to interpret
the model in terms of interactions between fields defined on sets of events.
E.g. a component $\alpha_{abc}$ can be regarded as a field variable
assigned to the set of events $\{ a, b, c \}$. There are $2^N$ field
variables corresponding to the number of possible subsets of the $N$ events.

This is only one step away from a field theory defined on string like
objects which pass through a sequence of events. Only the ordering is
missing.

Finally we note that the most basic representations of the Braid group
$B(N)$ are also defined to act on a space of dimension $2^N$. This suggests
that quantum group versions of event-symmetric field theories with $S(N)$
replaced by $B(N)$ might be possible.

The above observations were the original inspiration behind generalised
event-symmetric models involving string field theories and quantum groups.


\section*{Simplex Models}

In \cite{Gib95a} I constructed groups over a basis of discrete
strings in event-symmetric space-time.
Another class of groups closely related to the string groups is
based on sets of discrete events where the order does not matter
accept for a sign factor which changes according to the signature
of permutations,
\begin{equation}
(( a | b | c )) = -(( b | a | c ))
\end{equation}
\begin{equation}
	etc.
\end{equation}
A base element of length $n$ can be associated with a $n$-simplex
with vertices on the events in the element.

Single event simplices $(( a ))$ and a null simplex
$(( ))$ are included in the algebra.

Multiply by cancelling out any common events with appropriate sign factors.
To get the sign right, permute the events until the common ones are at the
end of the first set and at the start of the second in the opposite
sense. The elements can now be multiplied with the same rule as for the
open string. The same parity rules as for closed string apply. I.e. only
cancellations of an odd number of events is permitted.

The Lie product of two base elements can only be non-zero if they have an odd
number of events in common. e.g.
\begin{equation}
[ (( 1 | 2 | 3 )) (( 2 | 3 | 4 )) ]_{\pm} = 0
\end{equation}
\begin{equation}
[ (( 1 | 2 | 3 | 4 )) (( 4 | 3 | 2 | 5 )) ]_{\pm} = (( 1 | 5 ))
\end{equation}
This defines real and complex super-lie algebras which will be called
$simplex(0|N, {\Bbb R})$ and $simplex(0|N, {\Bbb C})$. These Lie algebras
are finite dimensional with dimension $2^N$.

An adjoint can be defined on the complex super-algebra in the usual way
\begin{equation}
                  \Xi = \sum \xi^C C
\end{equation}
\begin{equation}
           \Xi^{\dagger} = \sum \overline{\xi}^C i^{par(C)} C^T
\end{equation}
\begin{equation}
                              = \sum \overline{\xi}^C i^{len(C)} C
\end{equation}

If we take the sub-algebra of elements of $simplex(0|N, {\Bbb C})$ for which
\begin{equation}
                   \Xi^{\dagger} = - \Xi
\end{equation}
then this can be written in terms of their components as
\begin{equation}
                     \overline{\xi}^C = -i^{len(C)} \xi^C
\end{equation}
So
\begin{equation}
                    \xi^C = \phi^C i exp(i [\pi / 4] len(C))
\end{equation}
        With $\phi^C$ being real. If we use these as components writing,
\begin{equation}
                        \Xi = \sum \phi^C C_R
\end{equation}
\begin{equation}
                    C_R = i exp(- i [\pi / 4] len(C)) C
\end{equation}
It can be checked that the basis on $C_R$ has the same multiplication rules
as the basis on $C$ except for an extra minus sign when the number of
common events cancelled is 3 mod 4 just as in the algebra
$closed_{\pm}(0|N,{\Bbb R})$. This is the group $simplex(0|N)$.

The representations of these groups are families of fully antisymmetric
tensors. The Lie algebras are finite dimensional and it is therefore an
interesting exercise to determine how they correspond to the classification
of semi-simple Lie-algebras by factorising into well known compact groups.

An important remark about the simplex groups is that they have a resemblance
to the event-symmetric spinor models which can be seen when their components
are written as families of alternating tensors. In fact it is not
difficult to see that they are generated by the Clifford algebras for
$N$ dimensional space.

A matrix representation of the algebra can be constructed using Gamma matrices
which have size $2^{N/2} \times 2^{N/2}$ provided $n$ is even. In this
representation a mapping between the basic elements is defined by
\begin{equation}
(( a )) \rightarrow \gamma_a
\end{equation}
The gamma matrices satisfy the anticommutation relations,
\begin{equation}
            \gamma_a \gamma_b + \gamma_b \gamma_a = 2 \delta_{ab}
\end{equation}
The full algebra is generated from the linear span of all $2^N$ possible
products of the matrices e.g.
\begin{equation}
(( a | b )) \rightarrow \gamma_a \gamma_b
\end{equation}
The null simplex maps onto the identity matrix.

Since these are all linearly independent matrices with $2^N$ matrix elements
it follows that the algebra over the complex numbers is isomorphic to the
full matrix algebra $M(2^{N/2}, {\Bbb C})$. However, we are interested in
the $Z_2$ graded algebra where the parity is given by the size of the
simplex. It is possible to construct the gamma matrices so that they all have
elements in only the upper right and bottom left quadrants. The grading then
maps the algebra onto the super matrix algebra $M(L|L,{\Bbb C})$, where
$L = 2^{N/2-1}$. It follows that the Lie-superalgebra formed
{}from the graded anticommutators is just the super-symmetric affine
algebra and,
\begin{equation}
            simplex( N, {\Bbb C} ) \simeq gl( L|L, {\Bbb C} )
\end{equation}
while the adjoint defined on the signature algebra corresponds to the usual
adjoint on supermatrices so,
\begin{equation}
            simplex( N ) \simeq u( L|L )
\end{equation}
{}From this it is possible to construct and understand the invariants of the
algebra as invariants of the matrix super-groups. These are functions of the
supertrace of powers of the matrices.

The first order invariant turns out not to be the component corresponding
to the null simplex as you would expect. Instead it corresponds to the
simplex formed from all the $N$ events,
\begin{equation}
             U = (( 1,2,...,N ))
\end{equation}
This and higher order invariants seem to have anything but a local
nature since they are sums over products of simplices which include all
events but which have no event in common.

A second order invariant from the trace of the square is a sum over products
of two component tensors which have no event in common. This seems to be
just the opposite of what we want for a local theory but if we define a
relationship between a simplex and a dual simplex as follows
\begin{equation}
             \Xi^* = U\Xi
\end{equation}
Then the model is local.

It is interesting to compare this incomplete study of symmetries on
simplices with earlier work of a similar nature.
Finkelstein and Rodriguez also noted the importance of Clifford algebras in
this context \cite{Fin82,FiRo84,FiRo85}. The ideas presented here were derived
independently but the concurrence is important.
It is possible that the supersymmetry described here might lead to
further developments in this area.

There is an alternative interpretation of the simplex model which is
very instructive. The Clifford algebra can be compared directly with the
Fock space of a fermi gas (see e.g. \cite{Cor89}). The antisymmetric
tensors are then viewed
as antisymmetric wavefunctions describing fermion occupancy and the
dual mapping can be interpreted as a hole or anti-particle state.
A fermi gas is already a second quantised system in quantum field theory
and the quantisation procedure applied here is tantamount to a third
quatisation.


\section*{Duality}

The matrix models, when interpreted as event-symmetric, show quite clearly
how space-time symmetry and internal gauge symmetry could be unified. This
has always been regarded as the final goal which must be scored to unify
all physics, but why stop there? There are other symmetries which are often
overlooked. Many particle systems are invariant under exchange of identical
particles. The wavefunction is symmetric for bosons and anti-symmetric for
fermions. Since quantum field theory came to prominence this symmetry has
been demoted. It seen as a symmetry only of the quantum field and not a
true symmetry of the classical Lagrangian like the gauge symmetries.
Unification seems to be out of the question.

Is this conclusion justified? I would object. After all there is no classical
world, the $\hbar \rightarrow zero limit does not exist because changing \hbar$
only rescales our units of measurement. The universe is a quantum one and
invariance under particle exchange is as good a symmetry as any other.
Furthermore, the distinction between classical and quantum fields can
become blurred. This is dramatically demonstrated by the unity of dualities
in string theory \cite{HuTo94} which is apparently a duality between the
classical and quantum worlds \cite{Duf94}. This classical/quantum duality even
manifests itself even in the simple matrix models. Are the two
dimensional triangulated manifolds, which arises as the perturbation theory
of a matrix model, to be interpreted as the two dimensional classical
space-time of a 2-dimensional quantum gravity or the world sheet of a string
which are the Feynman diagrams of a $c < 1$ string theory? This kind of
duality where the Feynman diagrams of one model become the classical
configurations of another are quite common in event-symmetric models.

There is also an analogy between particle systems and event symmetric
systems which was exploited in the molecular models. It is now time to ask
if this could be more than just an analogy. Could there be a duality between
the symmetric group as it acts on space-time events and the symmetric group
acting on identical particles? The simplex model shows most clearly that
this is viable because it has a dual interpretation as a third quantisation
of a fermi gas aand an event-symmetric space-time model. Since string models
are likely to be quite closely related to this one, this greater unification
may be possible.


{\LARGE {\bf Acknowledgements}}

I would like to express my gratitude to those who have made the literature
accessible through multimedia on the internet. Special thanks must go to
Paul Ginsparg for setting up the physics e-print archives at Los Alomos.
Without the success of such a service my research would not have been
possible. I am also deeply indebted to the librarians at
SLAC, DESY and CERN for making the databases such as SPIRES database available
on the World Wide Web. My work has greatly
benefited from the facilities of the references and postscript databases.
I Thank John Baez for providing ``This Week's Finds in Mathematical Physics''
which have helped me catch up on some new physics.

A special thanks to the Biblioteque Interuniversaire Physique Recherche
at Jessieu for
permitting me free access to journals in more conventional form.
My gratitude is also extended to Eurocontrol for providing me with access
to the internet and allowing me to access it during my lunch breaks from
work.

\pagebreak


\end{document}